\documentclass[superscriptaddress,aps,nofootinbib,twocolumn]{revtex4}
\usepackage{graphicx}
\usepackage{amsmath,amssymb}

\def\bkR{{\rm I\kern-.17em R}}
\def\bkC{{\rm \kern.24em \vrule width.05em height1.4ex depth-.05ex \kern-.26em C}}
\def\to{\rightarrow}

\def\be{\beta}

\def\frac#1#2{{\textstyle{{#1}\over {#2}}}}

\def\lsim{\mathrel{\rlap{\lower4pt\hbox{\hskip1pt$\sim$}}
    \raise1pt\hbox{$<$}}}
\def\gsim{\mathrel{\rlap{\lower4pt\hbox{\hskip1pt$\sim$}}
    \raise1pt\hbox{$>$}}}
\def\sqr#1#2{{\vcenter{\vbox{\hrule height.#2pt
         \hbox{\vrule width.#2pt height#1pt \kern#1pt
         \vrule width.#2pt}
         \hrule height.#2pt}}}}

\def\laq{\raise 0.4 ex \hbox{$<$}\kern -0.8 em\lower 0.62 ex\hbox{$\sim$}}
\def\gaq{\raise 0.4 ex \hbox{$>$}\kern -0.7 em\lower 0.62 ex\hbox{$\sim$}}

\def\be{\begin{equation}}
\def\ee{\end{equation}}
\def\ba{\begin{eqnarray}}
\def\ea{\end{eqnarray}}

\def\dalemb#1#2{{\vbox{\hrule height.#2pt
        \hbox{\vrule width.#2pt height#1pt \kern#1pt \vrule width.#2pt}
        \hrule height.#2pt}}}

\def\dalemb#1#2{{\vbox{\hrule height.#2pt
        \hbox{\vrule width.#2pt height#1pt \kern#1pt \vrule width.#2pt}
        \hrule height.#2pt}}}

\def\gtorder{\mathrel{\raise.3ex\hbox{$>$}\mkern-14mu
             \lower0.6ex\hbox{$\sim$}}}
\def\ltorder{\mathrel{\raise.3ex\hbox{$<$}\mkern-14mu
             \lower0.6ex\hbox{$\sim$}}}

\begin{document}

\rightline{DF/IST-9.2009}
\rightline{December 2009}

\title{The singularity problem and phase-space noncanonical noncommutativity}

\author{Catarina Bastos\footnote{Also at Instituto de Plasmas e Fus\~ao Nuclear,IST. catarina.bastos@ist.utl.pt},
Orfeu Bertolami\footnote{Also at Instituto de Plasmas e Fus\~ao Nuclear,IST. orfeu@cosmos.ist.utl.pt}}

\vskip 0.3cm

\affiliation{Departamento de F\'\i sica, Instituto Superior T\'ecnico \\
Avenida Rovisco Pais 1, 1049-001 Lisboa, Portugal}

\author{Nuno Costa Dias\footnote{Also at Grupo de F\'{\i}sica Matem\'atica, UL, Avenida Prof. Gama Pinto 2, 1649-003, Lisboa, Portugal. ncdias@meo.pt}, Jo\~ao Nuno Prata\footnote{Also at Grupo de F\'{\i}sica Matem\'atica, UL, Avenida Prof. Gama Pinto 2, 1649-003, Lisboa, Portugal. joao.prata@mail.telepac.pt}}

\vskip 0.3cm

\affiliation{Departamento de Matem\'{a}tica, Universidade Lus\'ofona de
Humanidades e Tecnologias \\
Avenida Campo Grande, 376, 1749-024 Lisboa, Portugal}


\vskip 1cm

\begin{abstract}

\vskip 1cm

{The Wheeler-DeWitt equation arising from a Kantowski-Sachs model is considered for a Schwarzschild black hole under the assumption that the scale factors and the associated momenta satisfy a noncanonical noncommutative extension of the Heisenberg-Weyl algebra. An integral of motion is used to factorize the wave function into an oscillatory part and a function of a configuration space variable. The latter is shown to be normalizable using asymptotic arguments. It is then shown that on the hypersufaces of constant value of the argument of the wave function's oscillatory piece, the probability vanishes in the vicinity of the black hole singularity.}

\end{abstract}

\maketitle

\vskip 1cm


\section{Introduction}

Pioneering work in the context of quantum gravity and string theory \cite{CS} has led to great interest in noncommutativity aspects of quantum gravity, quantum field theory \cite{DS} and quantum mechanics \cite{BZA}. It is generally believed that some sort of noncommutativity may arise as one approaches the Planck scale. Noncommutativity may also be relevant in the context of black holes (BHs) physics. To avoid having to deal with an infinite number of degrees of freedom, one usually resorts to minisuperspace models. Starting from a suitable metric, one obtains via the ADM procedure the Wheeler-DeWitt (WDW) equation. Most often the solutions
of the latter are not square integrable and thus one faces the problem of determining a ``time" variable and a measure, such that on constant ``time" hypersurfaces, the wave-function is normalizable and the square of its modulus is a {\it bona fide} probability density. A square integrable wave function would allow, for instance, to address the problem of the BH singularity by computing the probability near the singularity .

Some steps in that direction were taken in Refs. \cite{Bastos2,Bastos5,Dominguez}. The Kantowski-Sachs (KS) metric was chosen, as it is integrable and it involves two scale factors, which is the minimum number of degrees of freedom for incorporating noncommutativity. Moreover, by a surjective mapping, one can describe the interior of the Schwarzshild black hole by the KS model. In Ref. \cite{Dominguez} only
configuration space noncommutativity was considered: $\left[\Omega, \beta \right] = i \theta$, where $\Omega, \beta$ are the scale factors and $\theta$ is a real constant. However, it is found in the cosmological and BH context, that this type of noncommutativity does not lead to any new qualitative features with respect to the commutative problem. On the other hand, it was shown in Refs. \cite{Bastos2,Bastos5}, that by imposing noncommutativity in the momentum sector as well, i.e. $\left[P_{\Omega}, P_{\beta} \right] = i \eta$, where $\eta$ is a real constant and $P_{\Omega}, P_{\beta}$ are momenta conjugate to $\Omega, \beta$ respectively, several interesting features arise: (i) For the Schwarzschild BH, it is found that the potential for the Schr\"odinger-like problem has a stable minimum. In the neighborhood of this minimum, one is able to perform a saddle point evaluation of the partition function and compute the relevant BH temperature and entropy; (ii) The solution of the WDW equation was shown to factorize into the product of an oscillatory function and a function which displays a conspicuous damping behavior. This damping does not lead to a full-fledged square integrable function, but the wave function does vanish at infinity, where the BH singularity is located. This leads one to conjecture whether another type of noncommutativity might yield a truly normalizable wave-function. In this letter it is shown that this is indeed possible. The crux of the argument lies on the choice of a noncanonical noncommutative algebra, which is nevertheless isomorphic to the Heisenberg-Weyl (HW) algebra. By noncanonical, it is meant that the commutation relations are not constants. Algebras of this kind have been thoroughly investigated in noncommutative quantum field theory (see, for instance, \cite{Daszkiewicz} and references therein) and in mathematical physics. For instance, the algebraic structure of the reduced phase space formulation of systems with second class constraints is typically a noncanonical noncommutative one \cite{Henneaux}.

In what follows, the KS metric formulation of the Schwarzschild BH is reviewed, and the noncommutative algebra and the isomorphic mapping to the HW algebra are defined. From this map, a quantum
representation of the noncanonical algebra is determined, which is then used to obtain the noncommutative WDW equation. Through a suitable constant of motion, the WDW equation is reduced into an ordinary differential equation. By resorting to asymptotic arguments and some results from the spectral theory of self-adjoint operators, it is shown that the solutions of this ordinary differential equation are square integrable.

\section{Phase-space noncommutative quantum cosmology}

In here the following system of units is used $\hbar = c =G=1$. The Schwarzschild metric for the interior region of a BH of mass $M$ is given by \cite{Bastos5}:
\ba\label{eq0.2}
ds^2=&-&\left({2M\over t}-1\right)^{-1}\!\!\!dt^2+\left({2M\over t}-1\right)dr^2+\nonumber\\
&+&t^2(d\theta^2+\sin^2\theta d\varphi^2)~.
\ea
This is an anisotropic metric, thus for $r<2M$ a Schwarzschild BH can be described as an anisotropic cosmological space-time. Indeed, the metric (\ref{eq0.2}) can be mapped by the KS metric
\cite{Bastos5,Dominguez}, which, in the Misner parametrization, can be written as
\be\label{eq0.3}
ds^2=-N^2dt^2+e^{2\sqrt{3}\beta}dr^2+e^{-2\sqrt{3}\beta}e^{-2\sqrt{3}\Omega}(d\theta^2+\sin^2{\theta}d\varphi^2)~,
\ee
where $\Omega$ and $\beta$ are scale factors, and $N$ is the lapse function. For $t<2M$, the identification
\ba\label{eq0.4}
N^2=\left({2M\over t}-1\right)^{-1} \!\!,\: e^{2\sqrt{3}\beta}&=&\left({2M\over t}-1\right)~,\nonumber\\
e^{-2\sqrt{3}\beta}e^{-2\sqrt{3}\Omega}&=&t^2~,
\ea
allows for turning the metric Eq. (\ref{eq0.3}) into the metric Eq. (\ref{eq0.2}). The black hole singularity lies 
at $t = r =0$, which according to Eqs. (\ref{eq0.4}) corresponds to $\beta, \Omega \longrightarrow +\infty$.

In the present work, one considers the following noncanonical extension of the HW algebra:
\ba\label{eq1.5}
\left[\hat{\Omega}, \hat{\beta} \right]\! &=&\! i \theta \left( 1 + \epsilon \theta \hat{\Omega} + {\epsilon \theta^2\over{1 + \sqrt{1- \xi}}} \hat{P}_{\beta} \right)\\
\left[\hat{P}_{\Omega}, \hat{P}_{\beta} \right]\! &=&\! i \left( \eta  + \epsilon (1 + \sqrt{1 - \xi})^2 \hat{\Omega} + \epsilon \theta (1 + \sqrt{1- \xi}) \hat{P}_{\beta} \right)\nonumber\\
\left[\hat{\Omega}, \hat{P}_{\Omega} \right]\! &=&\! \left[\hat{\beta}, \hat{P}_{\beta} \right]\! =\!i  \left( 1 + \epsilon \theta (1 + \sqrt{1- \xi}) \hat{\Omega} + \epsilon \theta^2 \hat{P}_{\beta} \right)\nonumber,
\ea
where $\theta$, $\eta$ and $\epsilon$ are positive constants and $\xi = \theta \eta <1$. The remaining commutation relations vanish. It is well known from Darboux's Theorem, that one can always find a system of coordinates, where locally the algebra looks like the HW algebra. For this algebra, this statement also holds globally, so that the algebra is isormorphic to the HW algebra. One calls this isomorphism a Darboux (D) transformation. D transformations are not unique. Indeed, the composition of a D transformation with a unitary transformation yields another D transformation.

Suppose that $\left(\hat{\Omega}_c, \hat{P}_{\Omega_c}, \hat{\beta_c}, \hat{P}_{\beta_c} \right)$ obey the HW algebra:
\ba\label{eq1.11}
\left[\hat{\Omega}_c, \hat{P}_{\Omega_c} \right] &=& \left[\hat{\beta}_c, \hat{P}_{\beta_c} \right] =i\nonumber\\
\left[\hat{\Omega}_c, \hat{\beta}_c \right] &=& \left[\hat{\Omega}_c, \hat{P}_{\beta_c} \right] =\left[\hat{\beta}_c, \hat{P}_{\Omega_c} \right] = 0~.
\ea
Then a suitable transformation would be:
\ba\label{eq1.7}
\hat{\Omega} \!\!&=&\!\! \lambda \hat{\Omega}_c -  {\theta\over2\lambda} \hat{P}_{\beta_c} + E \hat{\Omega}_c^2\hspace{0.2cm},\hspace{0.2cm}\hat{\beta} = \lambda \hat{\beta}_c + {\theta\over2 \lambda} \hat{P}_{\Omega_c} \nonumber\\
\hat{P}_{\Omega}\!\! &=&\!\! \mu \hat{P}_{\Omega_c} +  {\eta\over2\mu} \hat{\beta}_c \hspace{0.2cm},\hspace{0.2cm}\hat{P}_{\beta} = \mu \hat{P}_{\beta_c} -  {\eta\over2 \mu} \hat{\Omega}_c + F \hat{\Omega}_c^2.
\ea
Here, $\mu$, $ \lambda$ are real parameters such that $(\lambda \mu)^2 - \lambda \mu + \frac{\xi}{4} =0\Leftrightarrow 2 \lambda \mu = 1 + \sqrt{1- \xi}$, and
\ba\label{eq1.9}
&&E = - {\theta\over{1 + \sqrt{1- \xi}}} F\nonumber\\
&&F = - {\lambda\over\mu} \epsilon \sqrt{1- \xi} \left(1 + \sqrt{1- \xi} \right)~.
\ea
The inverse D is easily computed:
\ba\label{eq1.10}
\hat{\Omega}_c\!\! &=&\!\! {1\over\sqrt{1- \xi}}\!\! \left( \mu \hat{\Omega} + {\theta\over2 \lambda} \hat{P}_{\beta}  \right)\nonumber\\
\hat{P}_{\Omega_c}\!\! &=&\!\! {1\over\sqrt{1- \xi}} \!\! \left(\lambda \hat{P}_{\Omega} - {\eta\over2 \mu} \hat{\beta} \right)\nonumber\\
\hat{\beta}_c\!\! &=&\!\! {1\over\sqrt{1- \xi}}\!\! \left(\mu \hat{\beta} - {\theta\over2\lambda} \hat{P}_{\Omega} \right) \\
\hat{P}_{\beta_c}\!\!\! &= &\!\!\!\!{1\over\sqrt{1- \xi}}\!\! \left[\!\!\lambda \hat{P}_{\beta} \!\!+\!\!  {\eta\over2 \mu} \hat{\Omega} -\!\!{F \mu\over\sqrt{1- \xi}}\!\! \left( \!\!\hat{\Omega} + \!\!{\theta\over{1+\sqrt{1- \xi}}} \hat{P}_{\beta} \!\!\right)^2\right]\nonumber.
\ea
This means that these two algebras are isomorphic. It can be shown using Eqs. (\ref{eq1.7}) and (\ref{eq1.10}) that the algebra Eq. (\ref{eq1.5}) implies the HW algebra, Eq. (\ref{eq1.11}), and vice-versa.

One considers now the Hamiltonian associated to the WDW equation for the KS metric and a particular factor ordering \cite{Bastos2},
\be\label{eq1.12}
\hat H = - \hat{P}_{\Omega}^2 + \hat{P}_{\beta}^2 - 48 e^{-2 \sqrt{3} \hat{\Omega}}.
\ee
Upon substitution of Eqs. (\ref{eq1.7}), one obtains:
\ba\label{eq1.13}
\hat H&=& - \mu^2 \hat{P}_{\Omega_c}^2 - {\eta^2\over4 \mu^2} \hat{\beta}_c^2 - \eta \hat{\beta}_c \hat{P}_{\Omega_c} + \mu^2 \hat{P}_{\beta_c}^2 + {\eta^2\over4 \mu^2} \hat{\Omega}_c^2 +\nonumber\\
&+&F^2 \hat{\Omega}_c^4 - \eta \hat{\Omega}_c \hat{P}_{\beta_c} + 2 \mu F \hat{\Omega}_c^2 \hat{P}_{\beta_c} - {\eta F\over\mu} \hat{\Omega}_c^3 +\nonumber\\
&-&48 \exp \left( -2 \sqrt{3} \lambda \hat{\Omega}_c + {\sqrt{3} \theta\over\lambda}\hat{P}_{\beta_c}- 2 \sqrt{3} E  \hat{\Omega}_c^2 \right)~.
\ea
In the previous expression, the HW operators have the usual representation: $\hat{\Omega}_c = \Omega_c$, $\hat{\beta}_c = \beta_c$, $\hat{P}_{\Omega_c}= -i \frac{\partial}{\partial \Omega_c}$ and $\hat{P}_{\beta_c}= -i \frac{\partial}{\partial \beta_c}$. Let us now define the operator $\hat A$, which is analogous to the constant of motion that was defined for the problem discussed in Ref. \cite{Bastos2} (which considers the same spacetime setup, but assumes a canonical noncommutative algebra):
\be\label{eq1.14}
\hat A = \mu  \hat{P}_{\beta_c} + {\eta\over2 \mu} \hat{\Omega}_c.
\ee
A simple calculation reveals that for the noncommutative algebra Eq. (\ref{eq1.5}), $\hat A$ also commutes with the Hamiltonian $\hat H$, Eq. (\ref{eq1.12}), 
\be\label{eq1.15}
\left[\hat H, \hat A \right]=0~.
\ee 
Thus one looks for solutions $\psi \left( \Omega_c, \beta_c \right)$ of the WDW equation, 
\be\label{eq1.16}
\hat H \psi = 0~,
\ee
which are simultaneous eigenstates of $\hat A$. The most general solution of the eigenvalue equation $\hat A \psi = a \psi$ with $a$ real is,
\be\label{eq1.18}
\psi_a \left( \Omega_c, \beta_c \right)= {\cal R} \left( \Omega_c \right) \exp \left[ {i \beta_c\over\mu} \left(a - {\eta\over2 \mu} \Omega_c \right) \right],
\ee
where ${\cal R} \left(\Omega_c \right)$ is an arbitrary $C^2$ function of $\Omega_c$, which is assumed to be real.

From Eqs. (\ref{eq1.13}), (\ref{eq1.16}), and (\ref{eq1.18}), one gets after some algebraic manipulation
\ba\label{eq1.19}
&&\mu^2 {\cal R}''- 48 \exp \left( - {2 \sqrt{3}\over\mu} \Omega_c - 2 \sqrt{3} E \Omega_c^2 + {\sqrt{3} \theta a\over\mu \lambda} \right) {\cal R}+ \nonumber\\
&&- {2 \eta\over\mu} \left(\Omega_c+F\Omega_c^3\right){\cal R} + F^2 \Omega_c^4{\cal R} +\nonumber\\
&&+ a^2{\cal R} + \left({\eta^2\over\mu^2}+2 a F \right) \Omega_c^2{\cal R}=0~.
\ea
The dependence on $\beta_c$ has completely disappeared and one is effectively left with an ordinary differential equation for ${\cal R} \left(\Omega_c \right)$. Through the substitution, $\Omega_c = \mu z$, ${d^2\over d \Omega_c^2} = {1\over\mu^2} {d^2\over d z^2}$ and ${\cal R} \left( \Omega_c (z) \right) := \phi_a (z)$ one obtains a second order linear differential equation, which is a Schr\"odinger-like equation
\be\label{eq1.21}
- \phi_a'' (z) -V(z) \phi_a (z)= 0~,
\ee
where the potential function, V(z), reads:
\ba\label{eq1.22}
V(z) &=& - \left( \eta z -a \right)^2 - F^2 \mu^4 z^4 - 2  F \mu^2(\eta z-a) z^2 +\nonumber\\
&+& 48 \exp \left( -2 \sqrt{3} z -2 \sqrt{3} \mu^2 E z^2 + {\sqrt{3} \theta a \over\mu \lambda} \right).
\ea
Equation (\ref{eq1.21}) depends explicitly on the noncommutative parameters $\theta$, $\eta$, $\epsilon$ and the eigenvalue $a$.

Notice that $E>0$. One concludes from Eq. (\ref{eq1.22}) that asymptotically $(z \to \infty)$, the potential function is dominated by the term
\be\label{eq1.24}
V(z) \sim - F^2 \mu^4 z^4,
\ee
which leads to $L^2$ solutions of the Schr\"odinger equation (\ref{eq1.21}). Indeed, the Hamiltonian $H= - \frac{\partial^2}{\partial z^2} - F^2 \mu^4 z^4$ is self-adjoint \cite{Gitman} in $L^2$, its spectrum is continuous and the eigenfunction corresponding to the eigenvalue zero and eigenvalue $a$ of $\hat A$ has the asymptotic form:
 \be\label{eq1.25}
 \phi_a (z) \sim {1\over z} \exp \left[ \pm i \frac{F \mu^2}{3} z^3 \right]~.
 \ee
 Consequently, a generic solution of the WDW equation can be written as 
 \be\label{eq1.26}
 \psi (\Omega_c, \beta_c) = \int C(a) \psi_a (\Omega_c, \beta_c ) da~,
 \ee
 where $C(a)$ are arbitrary complex constants and $\psi_a (\Omega_c, \beta_c )$ is of the form:
 \be\label{eq1.27}
 \psi_a (\Omega_c, \beta_c ) = \phi_a \left(\frac{\Omega_c}{\mu} \right) \exp \left[\frac{i \beta_c}{\mu}\left(a - \frac{\eta}{2 \mu} \Omega_c \right) \right]~,
 \ee
 where $\phi_a(\Omega/\mu)$ is a the solution of Eq. (\ref{eq1.21}). Given that the wave function is oscillatory in $\beta_c$, it seems natural to fix a constant $\beta_c$-hypersurface. This suggests that a measure $\delta (\beta- \beta_c) d \beta d \Omega_c$ should be considered for the computation of probabilities. Thus, the BH probability $P(r=0,t=0)$ at the singularity would be given by:
 \ba\label{eq1.28}
&&P(r=0,t=0)= \lim_{\Omega_c, \beta_c \to + \infty} \int_{\Omega_c}^{+ \infty} | \psi(\tilde{\Omega}_c, \beta_c)|^2 d\beta_c d \tilde{\Omega}_c \nonumber\\
&&\simeq\lim_{\Omega_c \to + \infty} \int_{\Omega_c}^{+ \infty} \left|\phi_a \left({\tilde{\Omega}_c \over\mu} \right) \right|^2 d\tilde{\Omega}_c =0~,
 \ea
where in the last step, it has been used the fact that $\phi_a(\Omega/\mu)$ is square integrable.

\section{Conclusions}

In this work a KS cosmological model for the interior of a Schwarzshild BH is considered. It is shown that a noncanonical form of the phase-space noncommutativity leads to a natural factorization of the solutions of the WDW equation into an oscillatory function of the scale factor $\beta_c$ and a square integrable function of $\Omega_c$. On the constant $\beta_c$-hypersurfaces the solutions of the NCWDW equation display a zero asymptotic probability thus proving that in the vicinity of the singularity the BH probability distribution vanishes. This result seems to be quite robust as the asymptotic property is shared by all the solutions of the WDW equation, and is thus valid independently of the particular "initial" wave function.

The regularization of the BH singularity is a relevant novel result and relies on several steps. One first approaches the Schwarzschild BH through a map to the KS cosmological model. Noncanonical phase-space noncommutative relations are then imposed for the KS scale factors and the respective canonical conjugate momenta. The identification of a constant of motion of the associated classical problem leads to the identification of an operator $\hat{A}$, that commutes with the Hamiltonian of the system. Through the eigenvalue equation of $\hat A$ it is possible to reduce the problem of determining the solutions of WDW equation to that of solving a Schr\"odinger-like ordinary second-order differential equation, whose potential is asymptotically dominated by a quartic term. The corresponding Hamiltonian operator is self-adjoint in $L^2$ and the asymptotic behavior of its solutions admits a suitable probability definition at the singularity, which is shown to be vanishing.

\subsection*{Acknowledgments}

\vspace{0.3cm}

\noindent The work of CB is supported by FCT under the grant SFRH/BD/24058/2005. The work of NCD and JNP was partially supported by Grants No. PTDC/MAT/69635/2006 and PTDC/MAT/099880/2008 of the Portuguese Science Foundation FCT.



\begin{thebibliography}{99}

\bibitem{CS} A. Connes, M.R. Douglas and A. Schwarz, JHEP {\bf 9802} (1998) 003; N. Seiberg and E. Witten, JHEP {\bf 9909} (1999) 032.

\bibitem{DS} M.R. Douglas, N.A. Nekrasov, Rev. Mod. Phys. {\bf 73} (2001) 977; R. Szabo, Phys. Rep. {\bf 378} 207 (2003) 207.

\bibitem{BZA} O. Bertolami, J. G. Rosa, C. Arag\~ao, P. Castorina and D. Zappal\`a, Phys. Rev. {\bf D 72} (2005) 025010;  O. Bertolami, J. G. Rosa, C. Arag\~ao, P. Castorina and D. Zappal\`a, Mod. Phys. Lett. {\bf A 21} (2006) 795; Jian-zu Zhang, Phys. Rev. Lett. \textbf{93} (2004) 043002; Jian-zu Zhang, Phys. Lett. \textbf{B 584} (2004) 204; C. Acatrinei, Mod. Phys. Lett. {\bf A 20} (2005) 1437; C. Bastos, O. Bertolami, N.C. Dias and J.N. Prata, J. Math. Phys. {\bf 49} (2008) 072101; C. Bastos, O. Bertolami, N.C. Dias and J.N. Prata, Int. J. Mod. Phys. A {\bf 24} (2009) 2741.

\bibitem{Bastos2} C. Bastos, O. Bertolami, N.C. Dias and J.N. Prata, Phys. Rev. D {\bf 78} (2008) 023516.

\bibitem{Bastos5} C. Bastos, O. Bertolami, N.C. Dias and J.N. Prata, {\it Black Holes and Phase-Space Noncommutativity}, 0907.1818 [gr-qc], to appear in Phys. Rev. D.

\bibitem{Dominguez} J.C. L\'opez-Dominguez, O. Obreg\'on, M. Sabido and C. Ramirez, Phys. Rev. D {\bf 74} (2006) 084024.

\bibitem{Daszkiewicz} M. Daszkiewicz, J. Lukierski, M. Woronowicz, Phys. Rev. D {\bf 77} (2008) 105007.

\bibitem{Henneaux} M. Henneaux, C. Teitelboim, {\it Quantization of Gauge Fields} (Princeton University Press, 1992).

\bibitem{Gitman} B.L. Voronov, D.M. Gitman and I.V. Tyutin, {\it Self-adjoint differential operator associated with self-adjoint differential expressions}, quant-ph/0603187.



\end{thebibliography}
\end{document}